\documentclass[twocolumn,aps,prc,showpacs,superscriptaddress,floatfix]{revtex4}
\usepackage{graphicx}
\usepackage{dcolumn}
\usepackage{bm,ulem}
\usepackage{amsmath,amssymb}
\usepackage[usenames]{color}
\begin{document}
\title{Effect of neutron skin thickness on projectile fragmentation}
\author{Z. T. Dai}
\affiliation{Shanghai Institute of Applied Physics, Chinese Academy of Sciences, Shanghai 201800, China}
 \affiliation{University of Chinese Academy of Sciences, Beijing 100049, China}
\author{D. Q. Fang}\thanks{Email: dqfang@sinap.ac.cn}
\affiliation{Shanghai Institute of Applied Physics, Chinese Academy of Sciences, Shanghai 201800, China}
\author{Y. G. Ma}\thanks{Email: ygma@sinap.ac.cn}
\affiliation{Shanghai Institute of Applied Physics, Chinese Academy of Sciences, Shanghai 201800, China}
\affiliation{Shanghai Tech University, Shanghai 200031, China}
\author{X. G. Cao}
\affiliation{Shanghai Institute of Applied Physics, Chinese Academy of Sciences, Shanghai 201800, China}
\author{G. Q. Zhang}
\affiliation{Shanghai Institute of Applied Physics, Chinese Academy of Sciences, Shanghai 201800, China}
\author{W. Q. Shen}
\affiliation{Shanghai Institute of Applied Physics, Chinese Academy of Sciences, Shanghai 201800, China}

\date{\today}
\begin{abstract}
The fragment production in collisions of $^{48,50}$Ca+$^{12}$C at 50 MeV/nucleon are
simulated via the Isospin-Dependent Quantum Molecular Dynamics (IQMD) model followed
by the  {GEMINI code}.  {By changing the diffuseness parameter of neutron
density distribution to obtain different neutron skin size, the effects of
neutron skin thickness (${\delta}_{np}$) on projectile-like
fragments (PLF) are investigated. The sensitivity of isoscaling behavior to neutron skin size is studied, from which it is found that the isoscaling parameter $\alpha$ has a linear dependence on ${\delta}_{np}$. A linear dependence between ${\delta}_{np}$ and the mean $N/Z$ [N(Z) is neutron(proton) number] of PLF is obtained as well.}
The results show that thicker neutron skin will lead to smaller  {isoscaling parameter} $\alpha$ and N/Z. Therefore, it may be probable to extract information of neutron skin
thickness from  {isoscaling parameter} $\alpha$ and N/Z.
\end{abstract}

\maketitle
\section{Introduction}
The neutron skin of a nucleus has been one of the  {hottest} issues in
nuclear physics. It is defined as the difference
between the neutron and proton root-mean-square (RMS) radii:
${\delta}_{np}=\langle r_{n}^{2}\rangle^{1/2}-\langle
r_{p}^{2}\rangle^{1/2}$. The mechanism for the formation of neutron
skin can be expressed as follows. The proton-neutron interaction is
stronger than the proton-proton or neutron-neutron interaction.
Therefore, if the neutron number increases for neutron rich nuclei,
the mean potential for proton becomes deeper, while
the potential for neutron becomes shallower. Consequently, protons are
more deeply bound, but neutrons are more loosely bound, which will
form the neutron skin~\cite{skinform}.
The neutron skin thickness in neutron rich nuclei is crucial for studying
the equation of state (EOS) of asymmetric nuclear matter due to its strong
correlations with the symmetry energy, the slope
and curvature of symmetry energy at the saturation
density   {~\cite{reference03,reference07,reference08,reference09,reference10,reference11,reference12,reference13,reference14}.}
Thus neutron skin thickness and its effect in nuclear reactions
become an important research subject in nuclear physics.

 {Using} the Isospin-Dependent Quantum Molecular Dynamics (IQMD)
model, Sun \textit{et al}. have investigated the neutron to proton
ratio [$R(n/p)$] of emitted  {fragments} from projectile with different
neutron skin thickness and shown that there is a strong linear
 {correlation} between $R(n/p)$ and ${\delta}_{np}$, especially for
peripheral collisions~\cite{skinhalo1}.  Meanwhile, we have
investigated the correlation between the ratio of triton to $^{3}$He
[$R$(t/$^{3}$He)] and ${\delta}_{np}$, which exhibits the similar
linear  {relation}~\cite{symmetry1}. Projectile fragmentation is a
well-established technique for rare isotope productions by many
radioactive ion beam (RIB) facilities in the world. The projectile
fragmentation is used not only to study the reaction mechanism of
nuclear collision but also to extract some information on
fundamental physics~\cite{fragment_fang,fragment_ma,fragment_mocko}.
The production of heavy fragments will be affected by the neutron
and/or proton density distributions of the projectile nuclei. Thus
studies on the effect of neutron skin thickness over the
projectile-like fragment (PLF) will be very interesting.

In this paper, the relationship between ${\delta}_{np}$ and average $N/Z$,
the isotope distribution and the yield ratios of PLF will be studied within
the framework of IQMD model. The manuscript is organized as follows.
In Sec. II, we briefly describe the method, i.e. IQMD model plus
GEMINI  {code}.
In Sec. III,  {we present both the results and discussions.
Finally, a summary is given in Sec. IV.}

\section{The Framework Description}

\begin{figure}[t]
\resizebox{8.5cm}{!}{\includegraphics{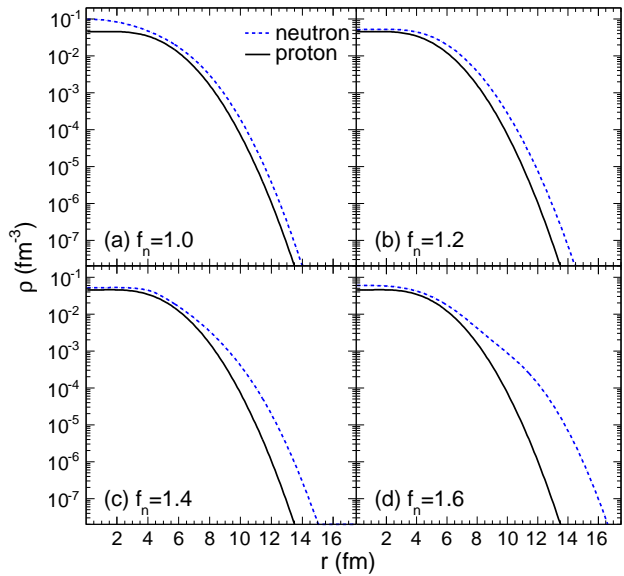}}
\caption{(Color online)  {The neutron and proton density profiles} calculated from
IQMD initialization for $^{50}$Ca at $t=0$ with
$f_{p}=1.0$ but different $f_{n}$.
(a) $f_{n}=1.0 (\delta_{np}=0.08$ fm); (b) $f_{n}=1.2 (\delta_{np}=0.21$ fm);
(c) $f_{n}=1.4 (\delta_{np}=0.37$ fm);
(d) $f_{n}=1.6 (\delta_{np}=0.54$ fm). } \label{dens_dis}
\end{figure}

\subsection{Dynamical model: IQMD}
The quantum-molecular-dynamics (QMD) approach is a many-body theory
to describe heavy ion collisions from intermediate to relativistic
energies. A general review about the model can be found in
 {Refs}.~\cite{reference38,reference103}. The IQMD model is based on the
QMD model with explicitly inclusion of isospin degrees of
freedom. It is well known that the dynamics at intermediate energies
is mainly governed by three parts: the mean field, two-body nucleon-nucleon
collisions and the Pauli blocking. Therefore, it is necessary to
include isospin degrees of freedom to the three parts for isospin
dependent dynamics. In initialization of the
projectile and target nuclei,  {neutrons and protons should be sampled separately in phase space}, especially for nuclei far  {away} from the $\beta$-stability line, of which the neutron and proton density
 distributions are much different. QMD model
has been widely and successfully used in heavy ion
collisions~\cite{skinhalo1,skinhalo2,skinhalo3,skinhalo4,Ma1995,fragment2,Ma-flow,other4,
Ma-vis,nst_qmd1,nst_qmd2,nst_qmd3,nst_qmd4}.
Details about the description of the IQMD model can be found in  {Refs}.~\cite{skinhalo1,symmetry1}.

In the present work, the following potential  {is} used,
\begin{eqnarray}
\lefteqn{U(\rho,\tau_z) =\alpha(\frac{\rho}{\rho_{0}})
+\beta(\frac{\rho}{\rho_{0}})^{\gamma} +\frac{1}{2}(1-\tau_z)V_c {}}
          \nonumber\\
&&\qquad\qquad {}+C_{sym}\frac{\rho_n-\rho_p}{\rho_0}\tau_z+U^{Yuk},
 \label{eq2}
\end{eqnarray}
 {where} $\rho_{0}=0.16$ fm$^{-3}$  {is} the normal nuclear matter density.
$\rho$, $\rho_{n}$, and $\rho_{p}$ are the total,
neutron, and proton densities, respectively. $\tau_z$ is the $z$-th
component of the isospin degree of freedom, which equals 1 or -1 for
neutrons or protons, respectively. The coefficients $\alpha$,
$\beta$ and $\gamma$ are parameters of the nuclear EOS. $C_{sym}$ is
the symmetry energy strength due to the difference between neutron
and proton asymmetry in nuclei. In
 {this work}, $\alpha =-356$ MeV, $\beta=303$ MeV and $\gamma =7/6$
are taken, which corresponds to the so-called soft EOS with
an incompressibility of $K = 200$ MeV and $C_{sym}=32$  {MeV}.
$V_c$ is the Coulomb potential and $U^{Yuk}$ is the Yukawa potential.  {Many theoretical studies show a strong correlation
 between symmetry energy and neutron skin size, while there is still large uncertainty of neutron skin size and symmetry energy~\cite{symm_skin1,symm_skin2}. As discussed in the reference\cite{symm_skin3}, both the symmetry and neutron skin have effects on the particle productions. Therefore, it is interesting to study the effects of neutron skin and symmetry energy separately. The correlations between symmetry energy and the particle or fragment productions have also been investigated in many theoretical studies\cite{symm_skin3,symm_frag1,symm_frag2}. The flexibility of adjusting independently the size of neutron skin of colliding nuclei is useful for our analyses in this work.
    Moreover, in this work, different neutron skin thicknesses mainly influence the neutron density distribution in the surface region of nuclei, so our study is focused on the peripheral collisions. In the peripheral collisions, the effect of neutron skin is larger than the effect due to the symmetry energy\cite{symm_skin3}. }

In the initialization of IQMD, the density distributions of
neutrons and protons are assumed to  {follow the} Fermi-type form
according to the droplet model ~\cite{reference39,reference40},
\begin{equation}
\rho_i(r)  =
\frac{\rho_{i}^{0}}{1+\exp(\frac{r-C_i}{f_{i}t_{i}/4.4})},\qquad
i=n,p,
 \label{eq3}
\end{equation}
where $\rho_{i}^{0}$ is the normalization constant which ensures
that the integration of the density distribution equals to the
number of neutrons ($i$=n) or protons ($i$=p); $t_i$ is the
diffuseness parameter; $C_i$ is the half density radius of neutron
or proton determined by the droplet model~\cite{reference40}.
\begin{equation}
C_{i} = R_{i}\left[ 1-(b_{i}/R_{i})^{2} \right], \qquad i=n,p,
\label{eq4}
\end{equation}
here $b_i = 0.413 f_{i}t_{i}$, $R_{i}$ is the equivalent sharp surface radius
of neutron or proton. $R_{i}$ and $t_{i}$ are given by the droplet model.
The factor $f_{i}$ is used to adjust the diffuseness of density distribution.
In this work, $f_{p}$ = 1.0 is used in Eq.(\ref{eq3}) for the proton density
distribution, while $f_{n}$ in Eq.(\ref{eq3}) is varied from 1.0 to 1.6 for
the neutron-rich projectile. Different values of $\delta_{np}$ can be deduced
from Eq.(\ref{eq3}) by changing $f_{n}$. Using the density distributions given
by the droplet model, initial  {coordinates} of nucleons in the nucleus are sampled
in terms of the Monte Carlo method. After initialization, the samples with satisfactory
stability and expected neutron skin size will be selected as candidates for collisions,
 {as described in Refs}.~\cite{skinhalo1,symmetry1}. Fig.~\ref{dens_dis} shows the neutron
and proton density profiles of $^{50}$Ca calculated from IQMD initialization.
In the simulation, $f_{n}$ is 1.0, 1.2, 1.4, 1.6 respectively. It can be seen that with
the increase of $f_{n}$, the neutron density distribution becomes more extended,
 {while that of proton is almost same in the four cases}.  {The related neutron skin thickness are 0.08 fm, 0.21 fm, 0.37 fm and 0.54 fm, respectively. The initial neutron skin can last a long time comparable with the reaction time, although a small-amplitude oscillation is visible. This level of stability of the initial samples is good enough for the purpose of this study.}
As a consequence, we can study the neutron skin effect on the production of fragments
with these samples. In this work, the fragments are constructed by a coalescence
model, in which nucleons with relative momentum smaller than P$_0$ = 300 MeV/c
and relative distance smaller than R$_0$ = 3.5 fm will be  {identified as} a cluster.

\subsection{Decay model: GEMINI}

\begin{figure}[htbp]
\resizebox{8.5cm}{!}{\includegraphics{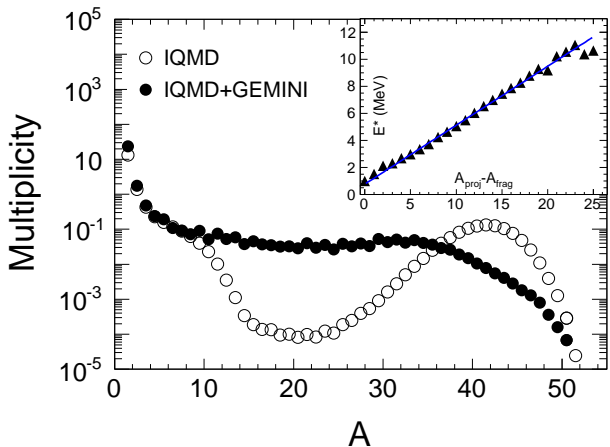}}
\caption{Fragment mass multiplicity distribution in 50MeV/nucleon
$^{50}$Ca+$^{12}$C with $f_{n}=1.0$ and $0.6<b/b_{\text{max}}<1.0$. Open circles represent IQMD calculation
at $t=200$ fm/c and solid points are results after the evaporation by GEMINI.
The excitation energy per nucleon of the prefragment
as a function of  {$A_{\text{proj}}-A_{\text{frag}}$}
calculated by IQMD simulations is shown in the inset.}
\label{multi_mc}
\end{figure}

\begin{figure}[t]
\resizebox{8.5cm}{!}{\includegraphics{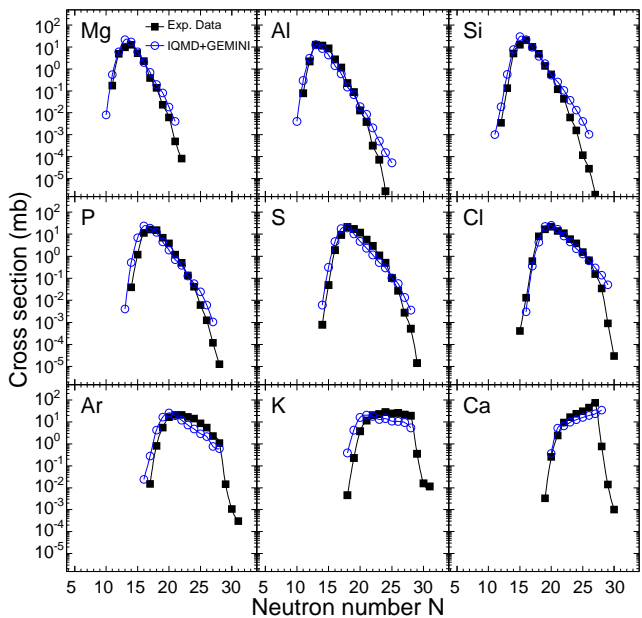}}
\caption{(Color online) The isotopic distributions plotted as a function of
neutron number N for elements with $12\leq Z \leq 20$ in 140 MeV/nucleon
$^{48}$Ca+$^{9}$Be.
Solid squares are the experimental data taken from Ref.~\cite{exp_data1},
and open circles are the calculated results with IQMD plus GEMINI.}
\label{exp_comp}
\end{figure}

The pre-fragments produced in IQMD are excited which are not comparable with
the final products measured experimentally, it is necessary to take into account
the evaporation effect  {to obtain} realistic results.
The GEMINI code is used to calculate de-excitation of these
fragments~\cite{reference_gemini}. GEMINI is a Monte Carlo code which allows not
just light-particle evaporation and symmetric fission, but all possible binary-decay modes.
It de-excites a source nucleus by a series of sequential binary decays until
the excitation energy of the excited fragment being unable to go further decay.
By using GEMINI, the fragment  {productions} are widely and successfully investigated
with QMD model~\cite{qmd_gemini1,qmd_gemini2,qmd_gemini4},
as well as, anti-symmetrized molecular dynamics (AMD) model~\cite{amd_gemini1,amd_gemini2}.
In IQMD, we can construct the  {angular} momentum and excitation
energy for each fragment which are used as input to GEMINI.
The  {angular} momentum of each nucleon in the fragment is calculated
according to classical  {mechanics}
\begin{equation}
\vec{L _{i}}=\vec{R _{i}}\times\vec{P _{i}},
 \label{eq5}
\end{equation}
where $\vec{R _{i}}$ and $\vec{P _{i}}$ are  {coordinate and momentum vector of the $i-$th nucleon of the fragment in the center of mass (CM) frame of the fragment}. The total  {angular} momentum
of the fragment is the summation of Eq.~(\ref{eq5}) over all nucleons in it.
The excitation energy of the primary fragment is calculated by following equation,
\begin{equation}
 {E^{*}=E^{\text{excited}}_{\text{bind}}-E^{\text{ground}}_{\text{bind}},}
 \label{eq6}
\end{equation}
where  {$E^{\text{excited}}_{\text{bind}}$} is the binding energy of the excited fragment calculated from IQMD,
and $E^{\text{ground}}_{\text{bind}}$ is the binding energy of the ground state taken from nuclear mass table~\cite{AUD03}.
The inset of Fig.~\ref{multi_mc} displays $E^{*}$ as a function of  {$A_{\text{proj}}-A_{\text{frag}}$},
where $A$ is mass number and the index proj and frag refer to the projectile and fragment, respectively.
 {Although the constructed excitation energy is model dependent, the excitation energy obtained from our IQMD calculation is comparable with that constructed from experimental data and model simulations~\cite{excited1,excited2,excited3}.}

With the calculated excitation energy and  {angular momentum, the fragments at $t=200$ fm/c
will be de-excited by using GEMINI. Since the main purpose of the
present work is to study the effect of neutron skin thickness of the
projectile on the production of heavy fragments, the calculations are focused on peripheral collisions. The reduced impact
parameter is used to describe the centrality of collision which is
defined as $b/b_{\text{max}}$, with $b_{\text{max}}$ being the
maximum impact parameter. For peripheral reaction, the
nucleon-nucleon interaction on the nuclear surface plays an
important role, where nucleons are loosely bound and the density
distributions are quite different for neutron and proton. Fig.~\ref{multi_mc} represents
 the comparison of mass versus multiplicity of fragments with and without GEMINI decay under the condition of $0.6<b/b_{\text{max}}<1.0$.
 In the peripheral collision of IQMD, the main reaction mechanism of fragment production is abrasion and evaporation of nucleons or light clusters, which leads to the excess of heavy products and deficiency of intermediate mass fragments (IMFs). In contrast, applying the afterburner significantly improves the fragments production since more complex fragments are emitted in the de-excitation calculation.}

\section{Results and Discussions}

\begin{figure*}[t]
\resizebox{17cm}{!}{\includegraphics{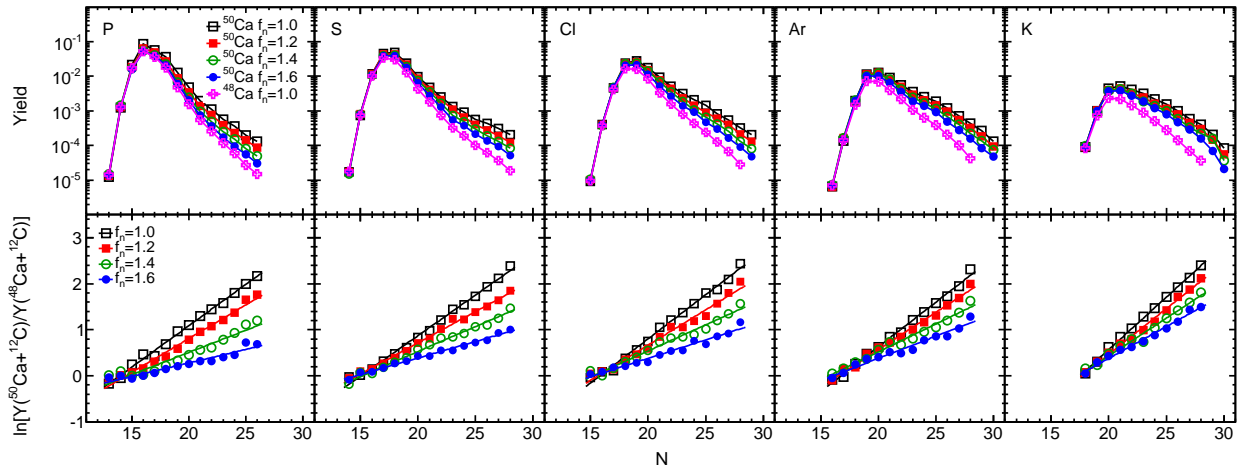}} \caption{(Color online) The isotopic yield per event
plotted as function of the neutron number $N$ with the charge number of fragment
varying from 15 to 19 in $^{50}$Ca+$^{12}$C and $^{48}$Ca+$^{12}$C at 50 MeV/nucleon (upper row).
 {The related isotopic yield ratios of the two reactions as function of $N$ is displayed in the lower row}. In the calculations, $f_{n}$ is varied from 1.0 to 1.6 for $^{50}$Ca+$^{12}$C,
while it is 1.0 for $^{48}$Ca+$^{12}$C.}
\label{iso_dis}
\end{figure*}

Before investigating the neutron skin effect in fragment production,
the isotopic distributions calculated by IQMD plus GEMINI are compared with
experimental data  {to evaluate the validity of the present method.}
In Fig.~\ref{exp_comp} the measured cross sections from Ref.~\cite{exp_data1}
and the calculated results of IQMD plus GEMINI for fragments with $12\leq Z \leq 20$
from $^{48}$Ca+$^{9}$Be at 140 MeV/nucleon are plotted.
The calculated yields are scaled to the experimental data for each isotope separately.
From this comparison, we can see that our calculations can reproduce the shape of
the isotopic distribution data quite well,
which suggests that the IQMD plus GEMINI model is reasonable for calculating
the heavy  {fragments}.
This is also demonstrated in  {Refs}.~\cite{qmd_gemini1,qmd_gemini4}.  {Meanwhile, Mock \textit{et al}.
have well reproduced this experimental data with heavy ion phase space exploration (HIPSE), Abrasion-Ablation (AA) and AMD~\cite{excited2,excited3}.}

As discussed in Ref.~\cite{nskin_frag}, the neutron skin plays  {a} significant role
in the production of fragment, especially for the production of neutron-rich nuclei.
Hence, to reveal the correlation between neutron skin and the fragment production
 {is interesting}.  Using the IQMD model, collisions of $^{48,50}$Ca projectile on
$^{12}$C target at 50 MeV/nucleon are simulated.  {In our simulations, the
 diffuseness parameter $f_{n}=1.0$ is used for $^{48}$Ca, while $f_{n}$ varies
from 1.0 to 1.6 for $^{50}$Ca. In the upper row of Fig.~\ref{iso_dis}, the production yield per event for five isotopes
 with the charge number varying from 15 to 19 are plotted. The reason why we choose the fragments with the charge number varying from 15 to 19 is explained as follows. The main purpose of this work is to study the neutron skin effects on fragments. In our IQMD, probability of fusion is small, so the fragments with charge number larger than projectile are neglected. The lighter fragments, which reach chemical balance at the end of the reaction will keep little information of the source nuclei, so the effect of neutron skin thickness on these fragments is very small. While the heavy residue can keep as much information of the projectile as possible. Therefore, fragments with the charge number varying from 15 to 19 are chosen as a probe of neutron skin.} Firstly, we compare the fragment
isotopic  {distributions} from reactions induced by projectiles with different neutron excess.
By comparing the two systems $^{48}$Ca+$^{12}$C and
$^{50}$Ca+$^{12}$C with $f_{n}=1.0$, it is demonstrated that there are more neutron-rich
fragments produced with increasing the neutron
excess of the projectile. This result is similar to that in
Ref.~\cite{fragment_fang}. Secondly, we compare the reactions
$^{50}$Ca+$^{12}$C with different $f_{n}$, which relates to different $\delta_{np}$. With the increase of
neutron skin thickness, yields of the isotope distributions will
decrease in the neutron-rich side but have almost no change in the
neutron-deficient side, which is consistent with the results  by statistical abrasion-ablation model~\cite{macw_saa}.

\begin{figure*}[t]
\resizebox{17.0cm}{!}{\includegraphics{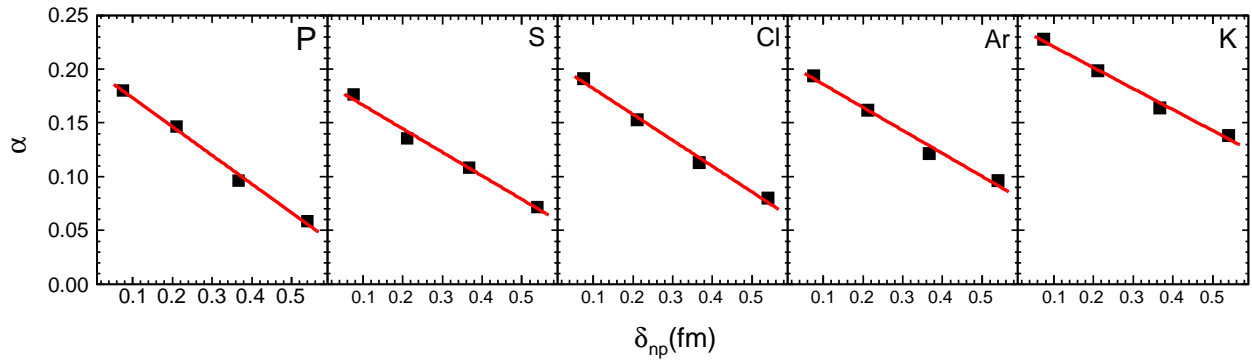}}
\caption{(Color online) The isoscaling parameter $\alpha$ as a function of $\delta_{np}$
for fragments with the charge number varying from 15 to 19 in 50 MeV/nucleon $^{50}$Ca+$^{12}$C.}
\label{scaling_skin}
\end{figure*}

\begin{figure}[b]
\resizebox{8.5cm}{!}{\includegraphics{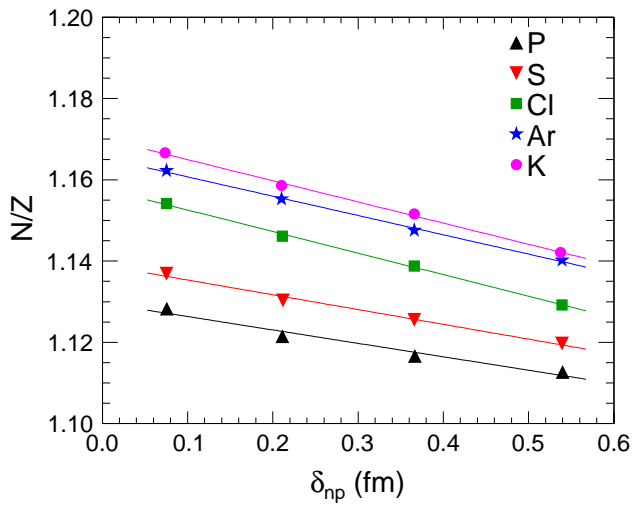}}
\caption{(Color online) The average $N/Z$ of fragment as a function of $\delta_{np}$ in
50 MeV/nucleon $^{50}$Ca+$^{12}$C. }
\label{ratio_nskin}
\end{figure}

Meanwhile, it has been shown that the yield ratios of isotopic fragments from two similar reactions
that differ only in the isospin asymmetry follow scaling
laws~\cite{scaling1,scaling2,scaling3,scaling4,scaling5,scaling6,scaling7}. The isotope yield ratios $R_{21}(N,Z)=Y_{2}(N,Z)/Y_{1}(N,Z)$
measured in two different nuclear reactions, 1 and 2, are found to exhibit an exponential relationship as
a function of the neutron number N and proton number Z~\cite{scaling1},
which can be expressed by following formula:
\begin{equation}
R_{21}(N,Z)=\frac{Y_{2}(N,Z)}{Y_{1}(N,Z)}=C\exp{(\alpha N+\beta Z)},
\label{eq7}
\end{equation}
where $\alpha$ and $\beta$ are two scaling parameters and $C$ is an
overall normalization constant.

The isoscaling phenomenon is
systematically investigated for fragments not heavier than $Z=8$. However,
it would be interesting to investigate the behavior of heavier fragments up to the region
of PLF, because the heavy residues, which is mainly produced by abrasion and
evaporating mechanism, may preserve some memory  of the source configuration,
for example, the neutron and proton density distribution. The isotopic scaling of heavy projectile residues is
 observed in both experiment and theoretical simulation~\cite{scal_resid1,scaling4,scaling5,scaling7,scaling8}.
 Consequently, from the research of  isoscaling of heavy residues of projectile,
 some information about neutron skin of neutron-rich projectile could be extracted,
 since the neutron skin does effect the production of fragments~\cite{nskin_frag}.  {This is our new start point.} In order to investigate the neutron skin size effect on isoscaling, the isotopic yield ratios
between two reactions $^{50}$Ca+$^{12}$C and $^{48}$Ca+$^{12}$C are plotted as a function
of the fragment neutron number, as shown in the lower row of Fig.~\ref{iso_dis}.
In the calculation,  $f_{n}=1.0$ is used for $^{48}$Ca, while $f_{n}=1.0,1.2,1.4$
and $1.6$ are used for $^{50}$Ca. From the figure, we can see that larger
neutron skin size will suppress the production of the neutron-rich fragment.
Although the neutron number and proton number of projectiles and targets are not changed
in reactions $^{50}$Ca+$^{12}$C and $^{48}$Ca+$^{12}$C, the neutron density of $^{50}$Ca
is different with different $f_{n}$. Eq.(\ref{eq7}) can be simplified for
isotopic yield ratio with the same charge number,
\begin{equation}
R_{21}(N)=C'\exp{(\alpha N)}.
\label{eq8}
\end{equation}
In the grand-canonical approximation, the scaling parameter $\alpha$ is equal to
the difference of the chemical potentials for neutrons in the two systems,
$\alpha=\mu_{n}/T$~\cite{scaling1,scaling2}.
 {The parameter $\alpha$ is extracted by
fitting $\ln[R_{21}(N,Z)]$ as a function of $N$. $\alpha$ dependence on $\delta_{np}$ for isotopes
with proton number varying from $Z=15$ to $Z=19$ is displayed in Fig.~\ref{scaling_skin}.} One can see that $\alpha$ decreases linearly
with the increasing of neutron skin thickness in the projectile.
It is indicated that the scaling parameter dose not keep constant for the heavy residues
with proton number close to the projectile, which is different from that for light fragment.
This different isoscaling behavior between heavy and light fragments may result from different  {formation} dynamics.
The light fragments are mainly produced from multi-fragmentation, in which the system has reached chemical balance,
which leads to the similar isoscaling behavior of light fragment.
While the heavy residues close to projectile originate from nucleons abrasion or evaporation.
In this case, most of the nucleon in projectile may only act as an spectator, and accordingly,
the system is not balanced. Nevertheless, the heavy residues will keep a memory of the projectile to some degree.
From this result some information of neutron skin could be extracted by measuring the isoscaling behavior of the heavy residues.

Finally, neutron skin effect on $N/Z$ of heavy PLF
is also investigated. The average $N/Z$ for PLF with charge number from $Z=15$ to $Z=19$
as a function of $\delta_{np}$ is plotted in Fig.~\ref{ratio_nskin}.
It displays that $N/Z$ decreases with the increasing
of $\delta_{np}$, and there is a good linear correlation between $N/Z$ and $\delta_{np}$.
This relationship could be regard as another probe for neutron skin thickness.
We also can see that for a certain $\delta_{np}$, the $N/Z$ will increase with the increasing
of charge number, which is similar to the results in Ref.~\cite{scal_resid1}

The neutron skin effect on fragment production, isoscaling, and $N/Z$ of residues could
be explained as following. For different neutron skin thickness, the mainly difference of
neutron density distributions is in the surface region. With increasing of $\delta_{np}$, more neutrons are
pushed to the surface of nuclei. Consequently, more neutrons will be abraded
in peripheral collision, which makes the residue less neutron-rich.
Large $\delta_{np}$ makes the neutron become more loosely bound, and more neutrons will be evaporated.

\section{Summary}
Using the isospin dependent quantum molecular (IQMD) model followed
by GEMINI, the neutron skin effect on the production of
projectile-like fragments are investigated in peripheral collisions of
50 MeV/nucleon $^{50}$Ca+$^{12}$C and $^{48}$Ca+$^{12}$C.
By changing the neutron diffuseness parameter of $^{50}$Ca
to obtain different neutron skin thickness ($\delta_{np}$), the dependence of
isotopic distributions, neutron to proton ratio ($N/Z$), and isoscaling behavior
of PLF on $\delta_{np}$ are studied.
It is demonstrated that larger $\delta_{np}$ suppresses the production of neutron-rich PLF.
This is because projectile with larger $\delta_{np}$ prefers to produce more free neutrons
and neutron-rich light clusters. The isoscaling behavior in
$^{50}$Ca+$^{12}$C and $^{48}$Ca+$^{12}$C are also investigated.
The extracted isoscaling parameter $\alpha$ decreases linearly with the increase
of $\delta_{np}$. The dependence of neutron to proton ratio ($N/Z$) of PLF on $\delta_{np}$
displays the similar trend. With the increase of $\delta_{np}$, more neutrons
will distribute in the nuclear surface.
Consequently, they will be abraded in peripheral collision more easily.
In conclusion, the isotopic distributions, isoscaling parameter $\alpha$,
and neutron to proton ratios $N/Z$ of PLF have dependence on $\delta_{np}$.
This dependence could probably be used to extract some information of neutron skin from experiments.

\section*{Acknowledgements}
This work was supported in part by the Major State Basic
Research Development Program in China under Contract No. 2013CB834405,
the National Natural Science Foundation of China under Contract No.s 11175231,
11035009, 11475244, 11421505 and 11205079 and
the Knowledge Innovation Project of Chinese Academy of Science 
under Grant No. KJCX2-EW-N01.

\end{document}